\documentclass[aps,prl,twocolumn,superscriptaddress,notitlepage]{revtex4-1}

\usepackage{graphicx}
\usepackage{amsmath}
\usepackage{bm}
\usepackage{color}
\usepackage{mdframed}
\usepackage[bookmarks=true,colorlinks=true,urlcolor=blue,linkcolor=blue,citecolor=blue,breaklinks]{hyperref}
\usepackage{amssymb}

\begin{document}

\title{Fermi Liquid Theory of $d$-Wave Altermagnets: Demon Modes and Fano-Demon States}

\author{Habib Rostami}
\email{hr745@bath.ac.uk}
\affiliation{Department of Physics, University of Bath, Claverton Down, Bath BA2 7AY, United Kingdom}

\author{Johannes Hofmann}
\email{johannes.hofmann@physics.gu.se}
\affiliation{Department of Physics, Gothenburg University, 41296 Gothenburg, Sweden}
\affiliation{Nordita, Stockholm University and KTH Royal Institute of Technology, 10691 Stockholm, Sweden}

\begin{abstract}
We develop a Fermi liquid theory of $d$-wave altermagnets and apply it to describe their collective excitation spectrum. We predict that in addition to a conventional undamped plasmon mode, where both spin components oscillate in phase, there is an acoustic plasmon (or {\em demon}) mode with out-of-phase spin dynamics. By analyzing the dynamical structure factor, we reveal a strong dependence of the demon's frequency and spectral weight both on the Landau parameters and on the direction of propagation. Notably, as a function of the propagation angle, we show that the acoustic mode evolves from a {\em hidden state}, which has zero spectral weight in the density excitation spectrum, to a weakly damped propagating demon mode and then (below a critical interaction parameter) to a {\em Fano-demon mixed state}, which is marked by a strong hybridization with particle-hole excitations and a corresponding asymmetric line shape in the structure factor. Our Letter paves the way for applications of altermagnetic materials in optospintronics by harnessing collective electron spin oscillations beyond traditional magnon spin waves.
\end{abstract}

\maketitle

\begin{figure}[t!]
    \centering
\includegraphics[width=0.9\linewidth]{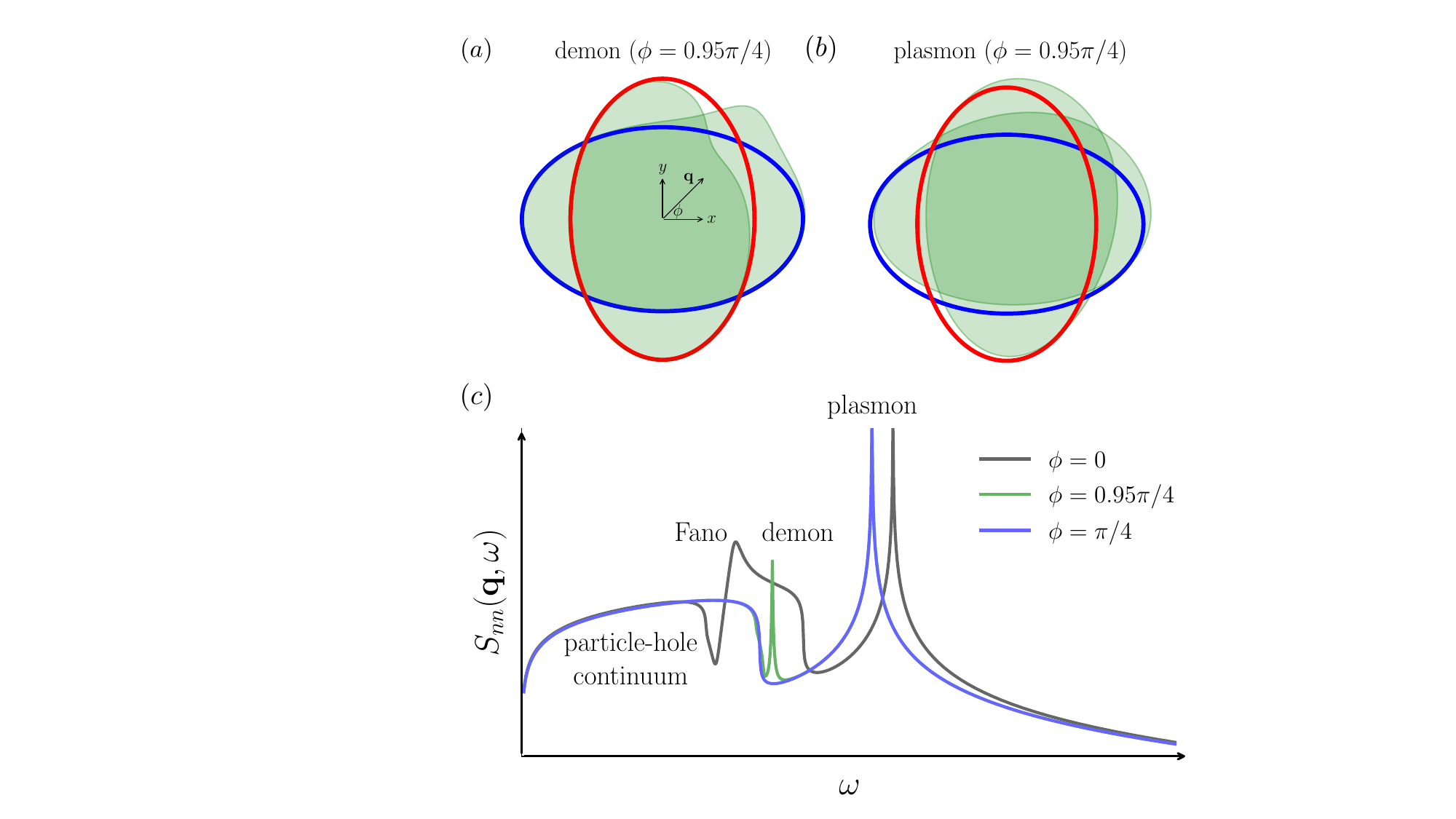}
\caption{
Microscopic Fermi surface deformations (green shaded areas) of (a)~the demon mode  and (b)~the plasmon mode for propagation near the nodal direction. The corresponding spin-$\uparrow$ and spin-$\downarrow$ Fermi surfaces are shown as blue and red ellipses, respectively. The demon mode involves an out-of-phase oscillation of both spin components, whereas for the plasmon mode both components oscillate in phase. (c)~Dynamical structure factor $S_{nn}({\bf q},\omega)$ as a function of frequency $\omega$ at fixed wave number \mbox{$q a_B = 0.4$} for three different directions of propagation \mbox{$\phi = 0, 0.95 \pi/4$}, and \mbox{$\pi/4$}. The plasmon is nearly isotropic, but there is a strong directional dependence of the demon mode contribution to the density fluctuations: For propagation exactly along the diagonal \mbox{$\phi=\pi/4$} (blue line), the demon has zero residue and is hidden in the density excitations. Near this nodal direction, the demon mode is weakly damped and present in the density excitations (green line). Finally, for propagation near the nematic axis (gray line), it hybridizes with the particle-hole continuum and forms a strongly asymmetric line shape, a Fano resonance.
}
\label{fig:1}
\end{figure}

{\em Introduction---} Altermagnetism refers to a novel magnetic configuration characterized by a compensated antiparallel ordering that breaks time-reversal symmetry and exhibits nonrelativistic spin splitting~\cite{Smejkal_PhysRevX011028,Smejkal_PhysRevX031042,Smejkal_PhysRevX040501,Bai_afm_2024,Krempasky2024,Zhu2024,Song2025}. Altermagnets possess spin-split bands, similar to ferromagnets, but exhibit no net magnetization, like antiferromagnets, thus representing a distinct magnetic phase that is characterized by a multipolar order parameter~\cite{McClarty_prl_2024} and results in unique phenomena such as the crystal Hall effect \cite{Libor_sciadv_2020}, an anomalous Hall effect \cite{Feng2022,Reichlova2024}, and the Nernst effect~\cite{han2024Nernst}. While the search for strictly two-dimensional (2D) altermagnets is ongoing (with proposed candidate materials, for example, V$_2$Te$_2$O, V$_2$Se$_2$O, and MnF$_2$~\cite{Ma2021,Brekke_prb_2023,Zeng_prb_2024,Liu_prl_2024,Milivojevic_2024}; see also Refs.~\cite{Smejkal_PhysRevX011028,Smejkal_PhysRevX031042,Smejkal_PhysRevX040501}), the recently realized layered altermagnets KV$_2$Se$_2$O~\cite{Jiang2025}, Rb$_{1-\delta}$V$_2$Te$_2$O~\cite{Zhang2025}, and Co$_{1/4}$NbSe$_2$~\cite{Alessandro2025} behave as quasi-2D systems that exhibit an effectively 2D nematic band dispersion with negligible out-of-plane dispersion. Here, opposite-spin sublattices are related by a spin-space rotation symmetry rather than translations or inversion~\cite{Smejkal_PhysRevX031042,McClarty_prl_2024}, which results in a nematic dispersion with a spin texture as depicted in Fig.~\ref{fig:1} for a $d$-wave altermagnet. The formation of such a nematic Fermi surface arises from a Pomeranchuk instability in the spin channel, leading to unconventional $d$-wave magnetic ordering \cite{Ahn_prb_2019,qian2025fragile,jungwirth2024supefluid} (where $g$-wave and $i$-wave altermagnet orders are also possible~\cite{Smejkal_PhysRevX031042,Ding_prl_2024,Zeng_adv_sci_2024,Yang2025}). This spin-nematic dispersion has been confirmed in several angle-resolved photoemission spectroscopy (ARPES) experiments~\cite{Krempasky2024,Zhu2024,Lee_prl_2024,Olena_sci_adv_2024,lin2024observation,Ding_prl_2024,Yang2025,Jiang2025}, although recent ARPES findings question the substantial spin splitting in RuO$_2$~\cite{Liu_Dawei_prl_2024}.

Distinct altermagnet orderings are reminiscent of systems in condensed matter physics, such as unconventional superconductors and Fermi liquid theory in superfluid He-3~\cite{jungwirth2024supefluid,Smejkal_PhysRevX040501}. The interplay of pronounced spin textures and anisotropic dispersion suggests that these materials will exhibit novel collective modes absent in conventional electron liquids. While recent research has investigated hydrodynamic transport~\cite{Herasymchuk_prb_2025}, Coulomb drag~\cite{Lin_prl_2025}, and Pomeranchuk instabilities~\cite{jungwirth2024supefluid,Smejkal_PhysRevX040501}, the development of a Landau Fermi liquid framework, especially concerning a spin-channel nematic Fermi surface, remains undeveloped in altermagnets.

In this Letter, we establish a Fermi liquid description for the nematic Fermi surfaces associated with $d$-wave altermagnets, where we use an elliptic coordinate parametrization that offers a natural way to define generalized nematic Landau parameters. We employ our framework to reveal the collective mode spectrum, which consists of  plasmon modes and  demon modes. The study of the demon mode~\cite{pines1956electron,DasSarma_prb_1981,DasSarma_prl_1998} is motivated by the recent experimental observation of  demon excitations in Sr$_2$RuO$_4$~\cite{Husain2023}. As illustrated in Figs.~\ref{fig:1}(a) and \ref{fig:1}(b), on a microscopic level, the demon and plasmon excitations are distinguished by out-of-phase and in-phase deformations of the spin-polarized Fermi surfaces, respectively. For the plasmon dispersion, we find at long wavelengths
\begin{equation}\label{eq:plasmon}
\hspace{-2.5mm}\omega_p=\sqrt{\frac{2\pi n e^2 q}{m(1\!-\!\alpha^2)}}\!\Bigg[1\!+\!\frac{q a_B(3\!+\!4F_0^s\!+\!\alpha^2(3\!+\!4F_0^a)\cos^2\!2\phi)}{8\sqrt{1\!-\!\alpha^2}}\Bigg]\!,
\end{equation}
where $n$ is the electron density, $m$ the isotropic effective mass, \mbox{$\alpha=|m_x-m_y|/(m_x+m_y)$} the anisotropy of the electron dispersion, and \mbox{$a_B=\hbar^2/me^2$} the Bohr radius. Equation~\eqref{eq:plasmon} takes an isotropic form with corrections at subleading order that depend on the propagation direction $\phi$ and the equation of state through the spin-symmetric and spin-antisymmetric isotropic Landau parameters \mbox{$F^s_0$} and \mbox{$F^a_0$}. For \mbox{$F^a_0 \geq 0$}, an acoustic demon mode exists with dispersion
\begin{equation}\label{eq:demon}
\omega_d = 
\frac{1 + 2 F_0^a}{\sqrt{1 + 4 F_0^a}} 
\, \frac{v_F q}{\sqrt{1-\alpha^2}},
\end{equation}
where $v_F$ is the isotropic Fermi velocity. While the dispersion~\eqref{eq:demon} is isotropic, we find that the contribution of the demon mode to the density fluctuations is strongly dependent on the direction of propagation. 
This is illustrated in Fig.~\ref{fig:1}(c), which shows the dynamical structure factor \mbox{$S_{nn}({\bf q},\omega)$} as a function of frequency $\omega$ for a fixed wave number $q$: our results indicate a nodal direction \mbox{$\phi=\pm\pi/4,\pm3\pi/4$} where the demon has zero residue and is hidden in the density excitations (blue line). In a finite range around this angle, given by (valid to leading order in $\alpha$)
\begin{align}
  \Delta \phi \simeq \arcsin \left[\frac{4 {F_0^a}^2}{\alpha(1+4F_0^a)}\right], %
\end{align}%
the demon mode is propagating and decoupled from the continuum of particle-hole excitations (green curve). Finally, for propagation closer to the nematic symmetry axis, the demon hybridizes with the particle-hole continuum, which is characterized by an asymmetric line shape (gray curve), a state that we call a {\em Fano-demon mixed state}~\cite{Fano_1961,Klein2020,Cappelluti_prb_2012}. Above a critical value \mbox{$F_0^a > F_{\mathrm{cr}} \approx (\alpha+\sqrt{\alpha+\alpha^2})/2$} (corresponding to \mbox{$\Delta \phi=\pi/2$}), the Fano-demon state disappears and the demon mode is propagating for all angles.

{\em Fermi liquid theory of $d$-wave altermagnets---}  The minimal model of a $d$-wave altermagnet has a dispersion  
\mbox{$\varepsilon_{\sigma}({\bf p}) = 2t [\cos p_x + \cos p_y] + \sigma t_J [\cos p_x - \cos p_y]$} with \mbox{$\sigma = \pm$} the electron spin index~\cite{Smejkal_PhysRevX040501}, which gives rise to  parabolic low-energy bands with anisotropic masses:
\begin{align}
    \varepsilon_{{\bf p},\sigma} =\frac{\hbar^2}{2m} \biggl(  \frac{p_x^2}{1+\sigma\alpha} + \frac{p_y^2}{1-\sigma\alpha} \biggr).
\end{align}
The Fermi surfaces \mbox{$\varepsilon_{{\bf p},\sigma}=\varepsilon_F$} for both spins are shown in blue (red) in Fig.~\ref{fig:1}(a).
We use an elliptic coordinate system with elliptic angle $\vartheta$ to parametrize the anisotropic Fermi surface and write the spin-up (\mbox{$\sigma=+$}) Fermi momentum as
\begin{align}
(p_x,p_y)  = 
   P_F (\cosh \xi_F \cos \vartheta, \sinh \xi_F \sin \vartheta)    
\end{align} 
with \mbox{$\xi_F  = \ln [(1+\sqrt{1-\alpha^2})/\alpha]^{1/2}$} and \mbox{$P_F = \sqrt{4\alpha m \varepsilon_F/\hbar^2}$}, while exchanging \mbox{$\sinh\xi_F\leftrightarrow\cosh\xi_F$} for the spin-down Fermi momentum (\mbox{$\sigma=-$}). The density of states for each spin in these elliptic coordinates is \mbox{${\cal D}_\sigma(\varepsilon_F,\vartheta) = \frac{m}{2 \pi\hbar^2} \sqrt{1-\alpha^2}$}, independent of both $\vartheta$ and spin. The Fermi velocity takes the form
\begin{align}
    {\bf v}_{F,\sigma} =
    v_F \left(\frac{\cos\vartheta}{\sqrt{1+\sigma\alpha}}, \frac{\sin\vartheta}{\sqrt{1-\sigma\alpha}}\right) \label{eq:vF}
\end{align}
with~\mbox{$v_F=\sqrt{2\varepsilon_F/{m}}$}, which resembles the expression for a circularly symmetric Fermi surface (note, however, that the elliptic angle $\vartheta$ is not a polar angle).
Our aim is to develop a Fermi liquid description for the dynamics of the quasiparticle distribution $f_{{\bf p},\sigma}$ on the nematic spin Fermi surfaces. In linearized form, this distribution follows the kinetic equation
\begin{align}
&\frac{\partial \delta f_{\bf p,\sigma}}{\partial t}
 + 
 {\bf v}_{F,\sigma}
 \cdot \frac{\partial\delta f_{\bf p,\sigma}}{\partial {\bf r} }
 -
 \frac{\partial\delta\varepsilon_{{\bf p},\sigma}}{\partial {\bf r}} 
 \cdot 
 \frac{\partial f_0(\varepsilon_{\bf p,\sigma})}{\partial {\bf p}} \nonumber \\
  &\qquad = {\cal J}_{\text{coll}}\{\delta f_{\bf p,\sigma}\} , \label{eq:boltzmann}
\end{align}
where \mbox{$\delta f_{\bf p,\sigma} = f_{{\bf p},\sigma} - f_0(\varepsilon_{\bf p,\sigma})$} is the deviation from global equilibrium, which is given by the Fermi-Dirac distribution \mbox{$f_0(\varepsilon_{\bf p,\sigma})=\Theta(\mu-\varepsilon_{\bf p,\sigma})$} with \mbox{$\mu=\varepsilon_F$},  
and \mbox{$\delta\varepsilon_{{\bf p},\sigma}$} describes Fermi liquid corrections to the quasiparticle energy,
\begin{align}
\delta\varepsilon_{\bf p,\sigma}  = \sum_{\bf p',\sigma'}  
{\cal F}^{\sigma\sigma'}_{{\bf p}{\bf p'}}   
    \delta f_{\bf p',\sigma'}  .
\end{align}
Here, ${\cal F}^{\sigma\sigma'}_{{\bf p}{\bf p'}}$ is the Landau function that parametrizes short-range interactions. Finally, ${\cal J}_{\text{coll}}$ is the collision integral that captures the collisional relaxation to local equilibrium.
\begin{figure*}
\centering   \includegraphics[width=\textwidth]{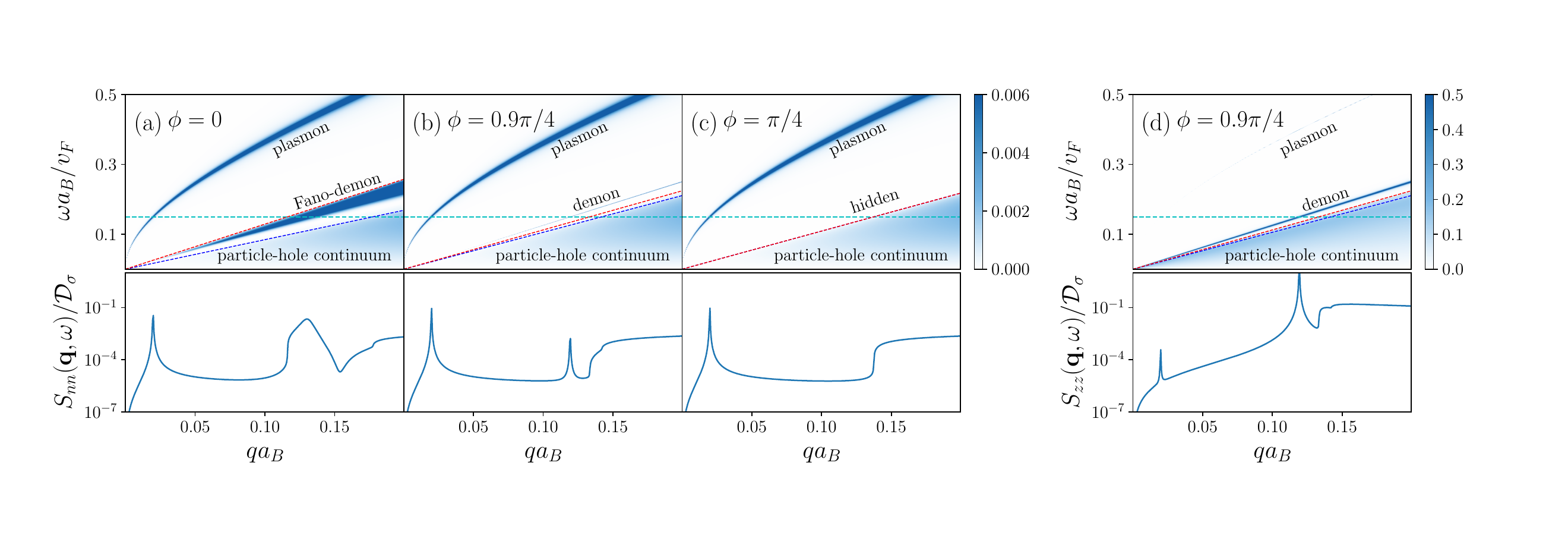}
\caption{
Dynamical density and spin structure factors $S_{nn}(\mathbf{q},\omega)$ and $S_{zz}(\mathbf{q},\omega)$ as a function of wave number $q$ and frequency $\omega$ for propagation directions (a)–(c) \mbox{$\phi=0,0.9\pi/4,\pi/4$} and (d) \mbox{$\phi=0.9\pi/4$}. Red and blue dashed lines 
mark the particle-hole continuum boundary for the two spin surfaces. The top row shows the spectral density and the bottom row shows cuts at a fixed energy \mbox{$\omega = 0.15 v_F/a_B$} (indicated by the dashed cyan line in the top panels). As $\phi$ increases from $0$ to $\pi/4$, the spectral weight of the demon branch becomes more concentrated, and the initially broad Fano-demon mode [panel (a)] sharpens into a well-defined Lorentzian peak [panel (b)]. 
At \mbox{$\phi = \pi/4$} [panel (c)], the spectral weight of the demon mode is suppressed in the density spectrum. By contrast, the spin structure factor (panel (d)) shows a strong demon peak and a faint plasmon peak. The parameters are \mbox{$\alpha = 0.4$}, \mbox{$\gamma = 10^{-4} v_F/a_B$}, \mbox{$F^{s}_{0} = 1$}, and \mbox{$F^{a}_{0} = 0.5$}.
}
\label{fig:2}
\end{figure*}
Collective-mode solutions of the Fermi liquid equation~\eqref{eq:boltzmann} take a plane-wave form with frequency~$\omega$ and wave vector~${\bf q}$,
\begin{equation}
    \delta f_{\bf p,\sigma} = \Bigl( - \frac{\partial f_0}{\partial \mu}\Bigr) e^{- i \omega t + i {\bf q} \cdot {\bf r}} \, h^{(\sigma)}(\vartheta) ,
\end{equation}
where we separate a nonanalytical prefactor and introduce the deformation function~$h$ of the Fermi surface. Solving Eq.~\eqref{eq:boltzmann} practically involves choosing appropriate basis functions for $h$. 
On a circular Fermi surface, the conventional choice is an expansion in polar angular harmonics, which form an orthonormal basis under the scalar product \mbox{$\langle \chi | \eta \rangle = \int_{\bf p} (-\partial f_0/\partial \mu) \chi^* \eta$} \cite{hofmann22}. This method breaks down for anisotropic Fermi surfaces, which couple different harmonic sectors. However, for $d$-wave Fermi surfaces, we find that a harmonic expansion with respect to the elliptic angle $\vartheta$ forms an orthonormal basis, which for the scalar product above follows directly from the definition of the density of states. Consequently, we represent the deviation function using the following elliptic harmonic expansion: 
\begin{equation}
    h^{(\sigma)}(\vartheta) =  \sum_{m} h^{(\sigma)}_{m}  e^{im\vartheta} . \label{eq:ellipticexpansion}
\end{equation}
This expansion is a key technical advance that allows for an efficient solution of the Fermi-liquid equations for a nematic Fermi surface. Equation~\eqref{eq:ellipticexpansion} suggests a modified parametrization of the Landau function in terms of elliptic Landau parameters by
\mbox{
${\cal F}^{\sigma\sigma'}_{\mathbf{p}\mathbf{p}'} = {\cal F}^{\sigma\sigma'}(\vartheta, \vartheta') = {\cal D}_\sigma^{-1}(\varepsilon_F) \sum_{m,m'} F^{\sigma\sigma'}_{mm'} e^{i m \vartheta - i m' \vartheta'}
$}, where we separate the density of states to define dimensionless Landau parameters $F^{\sigma\sigma'}_{mm'}$. We follow standard Fermi liquid convention to include only the leading ($s$-wave) Landau parameters, and, in the following analysis, we include  the leading ($s$-wave) Landau parameters,  \mbox{$F_{00}^{\uparrow\uparrow} = F_{00}^{\downarrow\downarrow} = F_0^s + F_0^a$} and \mbox{$F_{00}^{\uparrow\downarrow} = F_{00}^{\uparrow\downarrow} = F_0^s - F_0^a$}, which we parametrize in terms of the spin-symmetric and antisymmetric Landau parameters $F_0^s$ and $F_0^a$, respectively~\cite{pines_nozieres,baym2008,giuliani_vignale}. We have checked that including higher Landau parameters $F_{m\ge 1}^{s/a}$ has no qualitative effect on the demon and plasmon modes. In this parametrization, the long-range Coulomb interaction is taken into account in the Landau-Silin approach by separating a Hartree term \mbox{$F_{0}^s \to F_{0}^s + {\cal D}_\sigma(\varepsilon_F)v_q$} with \mbox{$v_q=2\pi e^2/q$} the Coulomb interaction in 2D. 

We adopt a generalized relaxation-time approximation for the collision integral, 
\mbox{$
{\cal J}_{\text{coll}}\{\delta f_{\mathbf{p},\sigma}\} 
= - (-\partial f_0/\partial \mu) \sum_m \gamma_m h^{(\sigma)}_m e^{im\vartheta}$}, with momentum-relaxing collisions with a constant decay rate in each elliptic angular channel except for the density deformation \mbox{$m=0$}, i.e., \mbox{$\gamma_m = \gamma \delta_{m\neq0}$}. 
The linearized Boltzmann equation for the collective modes projected onto the elliptic angular momentum basis then reads
\begin{align}\label{eq:FL_kinetic}
&(\omega+i\gamma_m) h^{(\sigma)}_m 
-\frac{q v_F}{2}  \sum_{\ell=\pm1}
u^{(\sigma)}_{\ell} h^{(\sigma)}_{m+\ell}
\\& 
- \frac{q v_F}{2} 
\sum_{\sigma',m',\ell=\pm1}  
u^{(\sigma)}_{\ell} \Bigl[\tilde v_q \delta_{m,-\ell}\delta_{m',0}+ F^{(\sigma\sigma')}_{m+\ell,m'} \Bigr]
h^{(\sigma')}_{m'} 
=0 \nonumber
\end{align}
with \mbox{$u^{(\sigma)}_{\pm} (\phi) =  \cos\phi /\sqrt{1+\sigma\alpha} \pm i \sin\phi/\sqrt{ 1-\sigma\alpha}$}, which depends on the polar angle $\phi$ of the wave vector $\bf q$, as well as \mbox{$\tilde v_q= {\cal D}_\sigma(\varepsilon_F) v_q= \sqrt{1-\alpha^2}/(a_B q
)$ }.

{\em Demon and plasmon collective modes---} 
Collective excitations of the $d$-wave altermagnet correspond to null vectors of the kinetic equation~\eqref{eq:FL_kinetic}, which describe the microscopic Fermi surface deformation of the mode, where the mode frequency follows from the roots of the coefficients matrix. Using the method of continued fractions~\cite{setiawan22,hofmann22,Rostami_prb_2025}, we reduce Eq.~\eqref{eq:FL_kinetic} to a \mbox{$6 \times 6$} coefficient matrix in the \mbox{$m = 0, \pm 1$} and spin \mbox{$\sigma = \pm$} subspace, where the coupling to higher harmonics is accounted for by the factor
\begin{align}\label{eq:X}
    X^{(\sigma)}_{\pm }(\phi) = \frac{h_{\pm2
}^{(\sigma)}}{h_{\pm1}^{(\sigma)}} = \frac{v_F q u^{(\sigma)}_{\mp} (\phi)}{\omega+i\gamma+\sqrt{(\omega+i\gamma)^2- |v_F q u^{(\sigma)}_{+} (\phi)|^2 }} .
\end{align}
The plasmon and demon dispersion discussed in this paper are obtained from a solution of the determinantal equation, which can be solved analytically in the long-wavelength limit to give Eqs.~\eqref{eq:plasmon} and~\eqref{eq:demon}, where the Fermi surface deformation illustrated in Fig.~\ref{fig:1} follows as \mbox{$h^{(\sigma)}(\vartheta) = h^{(\sigma)}_0 + \sum_{\ell=\pm} h^{(\sigma)}_\ell / [e^{-i\ell\vartheta} - X^{(\sigma)}_\ell(\phi)]$}
\footnote{The expression for \mbox{$h^{(\sigma)}(\vartheta) = \sum_m h^{(\sigma)}_m e^{im\vartheta} $} follows from \mbox{$h^{(\sigma)}_{\pm m} = [X^{(\sigma)}_{\pm}]^{m-1} h^{(\sigma)}_{\pm1}$}, leading to the expansion \mbox{$h^{(\sigma)}(\vartheta) = h^{(\sigma)}_0 + \sum_{\ell=\pm1}h^{(\sigma)}_\ell \{e^{i\ell\vartheta} +  \sum_{m\ge 2} [X^{(\sigma)}_{\ell}]^{m-1} e^{i\ell m\vartheta} \}$} that simplifies to the one used in the main text.}.

To uncover the spectral characteristics of collective modes we calculate the dynamical structure factor, which is related to the imaginary part of the particle density ($n$) and longitudinal current ($j_L$) correlation function $ \chi_{nj_L}$ through
\mbox{$
S_{nn}({\bf q},\omega) = -(q/\pi\omega) \text{Im}\chi_{nj_L}({\bf q},\omega)
$}, where \mbox{$ \delta n({\bf q},\omega) = \chi_{nj_L}({\bf q},\omega) A_L({\bf q},\omega) $} is the linear response to a longitudinal vector potential $A_L({\bf q},\omega)$. Accordingly, the kinetic equation~\eqref{eq:FL_kinetic} includes a source term on the right-hand side given by
\mbox{$
b^{(\sigma)}_m =(\omega v_F A_L/2) \sum_{\ell=\pm} \delta_{m,-\ell} u^{(\sigma)}_{\ell}(\phi) 
$}. Solving for the response field yields the dynamical structure factor 
\begin{align}\label{eq:defSqw}
    S_{nn}({\bf q},\omega) = -\frac{q}{\pi \omega A_L}
    {\rm Im}\sum_{\sigma}{\cal D}_{\sigma}(\varepsilon_F) h^{(\sigma)}_{0} .
\end{align}
The results shown in Fig.~\ref{fig:1}(c) are determined from Eq.~\eqref{eq:defSqw} with parameters \mbox{$q = 0.4/a_B$}, ellipticity \mbox{$\alpha = 0.4$}, and decay rate \mbox{$\gamma = 5 \times 10^{-4} v_F/a_B$}. Likewise, the spin-projected dynamical structure factor \mbox{$S_{zz}({\bf q},\omega)$} is obtained as the spin density in response to a spin-resolved source, i.e., by substituting \mbox{$b^{(\sigma)}_m \to \sigma b^{(\sigma)}_m$} and \mbox{$h^{(\sigma)}_0\to \sigma h^{(\sigma)}_0 $} ($\sigma=\pm1$).

{\em Hidden and Fano-demon modes---} To further examine how the spectral weight of the demon mode varies with the direction of propagation,  
Fig.~\ref{fig:2} shows the dynamical structure factor $S_{nn}$ as a function of frequency $\omega$ and momentum $q$ for the three angles \mbox{$\phi=0, 0.9\pi/4$, and $\pi/4$} [panels~(a)--(c)], as well as the spin-projected dynamical structure factor $S_{zz}$ for \mbox{$\phi=0.9\pi/4$} [panel~(d)].
The 2D plasmon is described by Eq.~\eqref{eq:plasmon} and mostly isotropic, whereas the spectral contribution of the acoustic demon mode shows a strong directional dependence. 
Starting from \mbox{$\phi = 0$} [Fig.~\ref{fig:2}(a)], the spectral response exhibits a broadened, asymmetric Fano line shape that becomes more pronounced with increasing $q$, characteristic of a Fano resonance. This asymmetric resonance arises from the interference between the demon mode and the particle-hole continuum. This behavior persists over a wide range of $\phi$.
As $\phi$ increases further, a transition occurs from the Fano-demon state to a coherent demon mode characterized by a well-defined Lorentzian peak [Fig.~\ref{fig:2}(b)]: here, we observe a sharp symmetric resonance with linear dispersion and notable spectral weight, indicating the emergence of a coherent demon pole outside the particle-hole continuum. Finally, at \mbox{$\phi = \pi/4$} [Fig.~\ref{fig:2}(c)], the acoustic mode disappears and becomes {\em hidden} in the density excitation spectrum. However, the demon peak remains at this angle in the spin structure factor $S_{zz}$ [Fig. \ref{fig:2}(d)], indicating a dominant out-of-phase density (or equivalently spin) oscillation and thus a finite spin moment, while the plasmon feature is weak, consistent with a negligible spin moment. This implies that the demon mode should couple to magnetic perturbations such as external magnetic fields or the internal magnetization.
\begin{figure}
\centering
\includegraphics[width=0.9\linewidth]{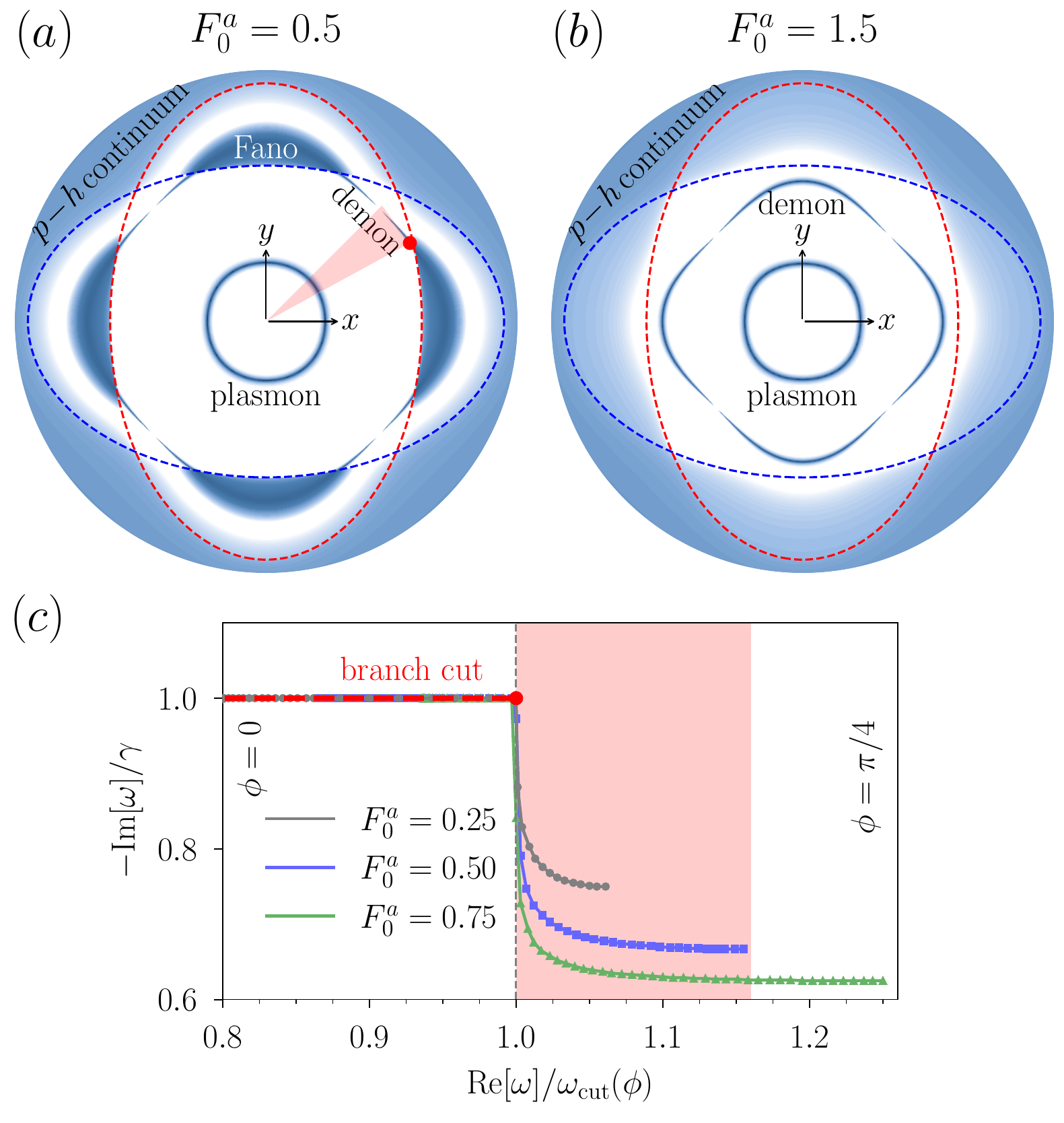}
\caption{Density plot of the dynamical structure factor \mbox{$ S_{nn}({\bf q}, \omega) $} in the wave-vector plane \mbox{$(q_x, q_y)$} for \mbox{$\omega = 0.4 v_F/a_B, \alpha = 0.4$,} and \mbox{$F_0^{s}=1$} for two different values of the Landau parameter (a)~\mbox{$F_0^{a}=0.5$} and (b)~\mbox{$F_0^{a}=1.5$}. Both panels show a weakly anisotropic plasmon mode (inner blue circle). Panel (a) shows a strongly anisotropic demon mode with (i)~a broad Fano–demon resonance region (intermediate blue spectral weight), (ii)~a narrow region with a demon excitation (red), and (iii)~a hidden branch along the nodal directions \mbox{$\phi=\pm\pi/4,\,\pm 3\pi/4$}. The transition point from the demon to the Fano–demon regime is marked by the red dot. In panel (b), the Fano–demon disappears and the demon mode lies entirely outside the particle–hole continuum with anisotropic dispersion. (c)~Trajectories of the Fano-demon and demon modes in the complex frequency plane as a function of propagation angle \mbox{$\phi \in [0, \pi/4]$} for three values of \mbox{$ F^a_0 = 0.25, 0.5$, and $0.75$}. The red dashed line marks the branch cut of particle-hole excitations, where the real part is rescaled. The red shaded area highlights the region between the Fano-demon and hidden modes (at \mbox{$\phi = \pi/4$}). 
}
    \label{fig:3}
\end{figure}

To further illustrate the transition from hidden to Fano-demon modes, Fig.~\ref{fig:3}(a) shows a density plot of the dynamical structure factor in the $q_x$-$q_y$ plane at fixed $\omega$. This visualization distinctly shows the marginal anisotropy of the plasmon mode (inner blue circle), compared to the significantly evident anisotropy of the demon collective mode. The plot exhibits three distinct features: ($i$) At \mbox{$\phi= \pm\pi/4, \pm 3\pi/4$}, the spectral weight disappears at the demon frequency, indicating the hidden mode. ($ii$) A true demon mode that is decoupled from the particle-hole continuum with finite spectral weight exists. ($iii$) A wide region depicts the Fano-demon state, where the demon mode merges with the particle-hole continuum, producing the Fano effect, evident as a non-Lorentzian asymmetric line shape with significant spectral weight. This Fano-demon state is not a genuine pole of the homogeneous kinetic equation. Figure~\ref{fig:3}(b) shows that once $F_0^{a}$ exceeds the critical value $F_{\rm cr}$, the demon-mode velocity increases and the mode is outside the particle–hole continuum for all angles (i.e., without a Fano-demon regime), while the spectral weight remains hidden in in the density excitations along the diagonals. This large-coupling regime is consistent with a recent RPA analysis with Coulomb interactions~\cite{gunnink25}, which reports a qualitatively similar angular dependence of the spin-demon mode in a $d$-wave altermagnet.

We also show in Fig.~\ref{fig:3}(c) the trajectory of the demon pole (as well as the Fano-demon quasipole) in the complex frequency plane with varying \mbox{$\phi\in[0,\pi/4]$}. Finding the demon pole trajectory in the complex plane requires finding the root of the homogeneous kinetic equation's coefficient matrix determinant, which is influenced by a branch cut due to the square root in the denominator of $X^{(\sigma)}_\pm$ [Eq.~\eqref{eq:X}]. As $\gamma$ approaches $0^+$, the branch cut's endpoint is determined by \mbox{$\omega_{\rm cut}(\phi) = {\rm max}[v_{F,+}, v_{F,-}] q$}, where for the two spins, we have \mbox{$v_{F,\sigma} = v_F \sqrt{(1-\sigma\alpha\cos2\phi)/({1-\alpha^2})}$}. At the critical onset $\phi$, indicated by the red dot in Fig.~\ref{fig:3}(a), the demon mode transitions into the Fano-demon state. For angles exceeding this onset value, a distinct demon pole is present in Fig.~\ref{fig:3}(c), even for \mbox{$\phi = \pi/4$}. However, for \mbox{$\phi = \pi/4$}, the demon spectral weight in Fig.~\ref{fig:3}(a) vanishes, indicating a hidden mode where a pole exists but the spectral weight vanishes. Conversely, for angles smaller than the onset value, the poles lie exactly on the branch cut, which is reminiscent of the mirage mode scenario discussed in isotropic Fermi liquids~\cite{Klein2020}. The demon mode on the branch cut then appears as a Fano resonance in the spectral weight shown in Fig.~\ref{fig:3}(a). Figure~\ref{fig:3}(c) also shows the demon pole for \mbox{$F^a_0=0.25, 0.5,\text{~and~}F^s_0=0.75$} with \mbox{$F^s_0=1$} and at \mbox{$\alpha=0.4$}. As \mbox{$F_0^a$} increases, the demon velocity in Eq.~\eqref{eq:demon} also increases, resulting in a greater separation from the particle-hole continuum and a shift away from the branch cut. This theoretical prediction is confirmed in Fig.~\ref{fig:3}(c), where for increasing $F^a_0$ the pole shifts further to the right of the branch point and reduces its imaginary part, indicating a sharper and longer-lived resonance. Conversely, smaller $F^a_0$ values cause the demon pole to approach the branch cut, broaden, and increase its imaginary component.

In summary, we study collective modes in $d$-wave altermagnets. In addition to a nearly isotropic plasmon mode, we predict an emergent demon mode whose properties strongly depend on the propagation direction and Landau parameters, with notable spectral weight in Fano-demon states. Beyond using high-resolution scanning-probe spectroscopy to isolate a single crystalline domain, the Fano–demon's strong and broad spectral feature makes disorder and domain wall effects only a subleading concern even under angle averaging across randomly oriented domains. Moreover, once the interaction exceeds the critical threshold (\mbox{$F_0^a>F_{\rm cr}$}), the demon mode persists over nearly all directions, ensuring observability in polydomain samples.
With spin–orbit coupling (SOC), there can be a gap opening along the diagonals that splits the Fermi surfaces into spin-mixed Fermi pockets. We expect that this can suppress the demon mode along the diagonals while for all other directions both the demon and the Fano-demon persist, albeit with a SOC-modified spin moment and angular width.
Candidate materials where our prediction could be observed include layered altermagnets such as $\mathrm{Rb}_{1-\delta}\mathrm{V}_2\mathrm{Te}_2\mathrm{O}$. 
To estimate the demon and Fano-demon velocities in these compounds, we can compare with recent ARPES data for the spin-up and spin-down Fermi pockets in Ref.~\cite{Zhang2025}. These experiments obtain Fermi velocities of 
\mbox{$v_{F,\uparrow}\!\approx\!4\times10^5~\mathrm{m/s}$} and \mbox{$v_{F,\downarrow}\!\approx\!6.7\times10^5~\mathrm{m/s}$} along the \mbox{$\phi=0$} direction, which by Eq.~\eqref{eq:demon} set the scale of the demon velocity. This study paves the way for novel applications of altermagnetic materials in optospintronics by utilizing demon mode spin oscillations in addition to conventional magnon spin waves.

\begin{acknowledgments}
{\em Acknowledgments---}We thank Viktor Fril\'en for discussions. This work is supported by the Engineering and Physical Sciences Research Council (Grant No. UKRI122), Royal Society (Grant No.\ IES\textbackslash R2\textbackslash 242309),  Vetenskapsr\aa det (Grants No.~2020-04239 and No.~2024-04485), the Olle Engkvist Foundation (Grant No.~233-0339), the Knut and Alice Wallenberg Foundation (Grant No.~KAW 2024.0129), and Nordita.
\end{acknowledgments}

{\em Data availability–}The data that support the findings of this Letter are openly available \cite{dataset}.

\bibliography{bib_demon}

\end{document}